\documentclass[twocolumn,
prb
,superscriptaddress
,floatfix
,aps
,10pt
,showpacs
]{revtex4-1}

\usepackage{graphicx}
\usepackage{amssymb}
\usepackage{amsmath}
\usepackage[usenames,dvipsnames]{color}
\usepackage{comment}

\newcommand{\vect}[1]{{\mathbf #1}}
\newcommand{\vectgr}[1]{{\boldsymbol#1}}    
\newcommand{\Frac}[2]{\displaystyle\frac{#1}{#2}}

\begin{document}


\title{On multicomponent polariton superfluidity in the optical
  parametric oscillator regime}

\author{A. C. Berceanu}
\affiliation{Departamento de F\'isica Te\'orica de la Materia
  Condensada \& Condensed Matter Physics Center (IFIMAC), Universidad
  Aut\'onoma de Madrid, Madrid 28049, Spain}

\author{L. Dominici} 
\email{lorenzo.dominici@gmail.com} 
\affiliation{NNL, Istituto Nanoscienze-CNR, Via Arnesano, 73100
  Lecce, Italy}
\affiliation{Istituto Italiano di Tecnologia, IIT-Lecce, Via Barsanti,
73010 Lecce, Italy}

\author{I. Carusotto} 
\affiliation{INO-CNR BEC Center and Universit\'a di Trento, via
  Sommarive 14, I-38123 Povo, Italy}

\author{D. Ballarini} 
\affiliation{NNL, Istituto Nanoscienze-CNR, Via Arnesano, 73100
  Lecce, Italy}
\affiliation{Istituto Italiano di Tecnologia, IIT-Lecce, Via Barsanti,
73010 Lecce, Italy}

\author{E. Cancellieri} 
\affiliation{Department of Physics and Astronomy, University of
  Sheffield, Sheffield, S3 7RH, UK}

\author{G. Gigli}
\affiliation{NNL, Istituto Nanoscienze-CNR, Via Arnesano, 73100
  Lecce, Italy}

\author{M. H. Szyma\'nska}
\affiliation{Department of Physics and Astronomy, University College
  London, Gower Street, London, WC1E 6BT, UK}

\author{D. Sanvitto} 
\affiliation{NNL, Istituto Nanoscienze-CNR, Via Arnesano, 73100
  Lecce, Italy}

\author{F. M. Marchetti} 
\email[Corresponding author: ]{francesca.marchetti@uam.es} 
\affiliation{Departamento de F\'isica Te\'orica de la Materia
  Condensada \& Condensed Matter Physics Center (IFIMAC), Universidad
  Aut\'onoma de Madrid, Madrid 28049, Spain}

\date{March 12, 2015}       

\begin{abstract}
  Superfluidity, the ability of a liquid or gas to flow with zero
  viscosity, is one of the most remarkable implications of collective
  quantum coherence. In equilibrium systems like liquid ${}^4$He and
  ultracold atomic gases, superfluid behaviour conjugates diverse yet
  related phenomena, such as persistency of metastable flow in
  multiply connected geometries and the existence of a critical
  velocity for frictionless flow when hitting a static defect.
  The link between these different aspects of superfluid behaviour is
  far less clear in driven-dissipative systems displaying collective
  coherence, such as microcavity polaritons, which raises important
  questions about their concurrency.
  With a joint theoretical and experimental study, we show that the
  scenario is particularly rich for polaritons driven in a three-fluid
  collective coherent regime so-called optical parametric oscillator.
  On the one hand, the spontaneous macroscopic coherence following the
  phase locking of the signal and idler fluids has been shown to be
  responsible for their simultaneous quantized flow metastability.
  On the other hand, we show here that pump, signal and idler have
  distinct responses when hitting a static defect; while the signal
  displays hardly appreciable modulations, the ones appearing in pump
  and idler are determined by their mutual coupling due to nonlinear
  and parametric processes.
\end{abstract}

\pacs{03.75.Kk, 71.36.+c, 42.65.Yj}






\maketitle
  
\section{Introduction}
Microcavity polaritons, the novel quasiparticles resulting from the
coherent strong coupling between quantum well excitons and cavity
photons~\cite{kavokin_laussy}, have unique mixed matter-light
properties that none of their constituents displays on its
own. Because of their energy dispersion and their strong non-linearity
inherited from the excitonic components, polaritons continuously
injected by an external laser into a pump state with suitable
wavevector and energy can undergo coherent stimulated scattering into
two conjugate states~\cite{ciuti00:prb,ciuti01,ciuti03}, the signal
and the idler, a process known as optical parametric oscillator (OPO).
Since their first
realisation~\cite{stevenson00,savvidis00:prl,savvidis00:prb,baumberg00:prb,saba01},
the interest in microcavity optical parametric phenomena has involved
several fields of fundamental and applicative
research~\cite{edamatsu2004,savasta05:prl,lanco2006,abbarchi2011,ardizzone2012,xie2012,lecomte2013}.

Recently, considerable resources have been invested in exploring the
fundamental properties of parametric processes, including the
possibility of macroscopic phase coherence and superfluid
behaviour~\cite{Carusotto2013a}.
In spite of the coherent nature of the driving laser pump, the OPO
process belongs to the class of non-equilibrium phase transitions, in
which a $U(1)$ phase symmetry is spontaneously
broken~\cite{wouters06b}.
While the phase of the pumped mode is locked to the incident laser,
the ones of signal and idler are free to be simultaneously rotated in
opposite directions.
Because of this phase freedom, recent experiments~\cite{sanvitto10}
have tested the OPO superfluid properties by exploring the physics of
the signal-idler order parameter, demonstrating the existence and
metastability of vortex configurations. As the order parameter
involves both signal and idler, their phase winding have opposite
signs~\cite{sanvitto10,marchetti10,marchetti_review}.  Crucially, this
makes both OPO fluids to display quantized flow metastability
simultaneously.

While in equilibrium condensates different aspects of superfluidity
are typically closely related~\cite{LeggettRMP_99}, this is no longer
true in a non-equilibrium context as for microcavity
polaritons~\cite{Carusotto2013a}.
In particular, those aspects of superfluidity related to the
frictionless flow around defects are expected to be much more involved
in OPO condensates than for any other investigated polariton
condensates, such as for the case of incoherent
pumping~\cite{kasprzak06:nature,wouters2010}, and single-state
resonantly pumped microcavities~\cite{amo09_b}.
Independently on the pumping scheme, the driving and the polariton
finite lifetime prompt questions about the meaning of superfluid
behaviour, when the spectrum of collective excitations is complex
rather than real, raising conceptual interrogatives about the
applicability of a Landau criterion~\cite{wouters2010}.
Yet, an additional complexity characterises the OPO regime, where the
simultaneous presence of three oscillation frequencies and momenta for
pump, signal and idler correspondingly multiplies the number of
collective excitation branches~\cite{wouters06b}.
Note that from the experimental point of view, pioneering
experiments~\cite{amo09} have observed a ballistic non-spreading
propagation of signal/idler polariton wavepackets in a triggered-OPO
configuration.
However, given the complexity of the dynamics as well as the nonlinear
interactions involved in this time-dependent
configuration~\cite{szymanska2010}, theoretical understanding of these
observations is not complete yet.

This Article reports a joint theoretical and experimental study of an
OPO configuration where a wide and steady-state condensate hits a
stationary localised defect in the microcavity.
Contrary to the criterion for quantized flow metastability for which
signal and idler display simultaneous locked responses, we find that
their scattering properties when the OPO hits a static defect are
different.
In particular we investigate the scattering properties of all three
fluids, pump, signal and idler, in both real and momentum space. We
find that the modulations generated by the defect in each fluid are
determined not only by its associated Rayleigh scattering ring, but
each component displays additional rings because of the cross-talk
with the other components imposed by nonlinear and parametric
processes.
We single out three factors determining which one of these rings
influences the most each fluid response: the coupling strength between
the three OPO states, the resonance of the ring with the blue-shifted
lower polariton dispersion, the values of each fluid group velocity
and lifetime together establishing how far each modulation can
propagate from the defect.
The concurrence of these effects implies that the idler strongly
scatters inheriting the same modulations as the pump, while the
modulations due to its own ring can propagate only very close to the
defect and cannot be appreciated. Yet, the modulations in the signal
are strongly suppressed, and not at all visible in experiments,
because the slope of the polariton dispersion in its low momentum
component brings all Rayleigh rings coming from pump and idler out of
resonance.

Note that the kinematic conditions for OPO are incompatible with the
pump and idler being in the subsonic regime. Thus, the coupling
between the three components always implies some degree of scattering
in the signal. In practice, the small value of the signal momentum
strongly suppresses its actually visible modulations, something
confirmed by the experimental observations.

\begin{figure}[h!]
\centering
\includegraphics[width=1.0\columnwidth]{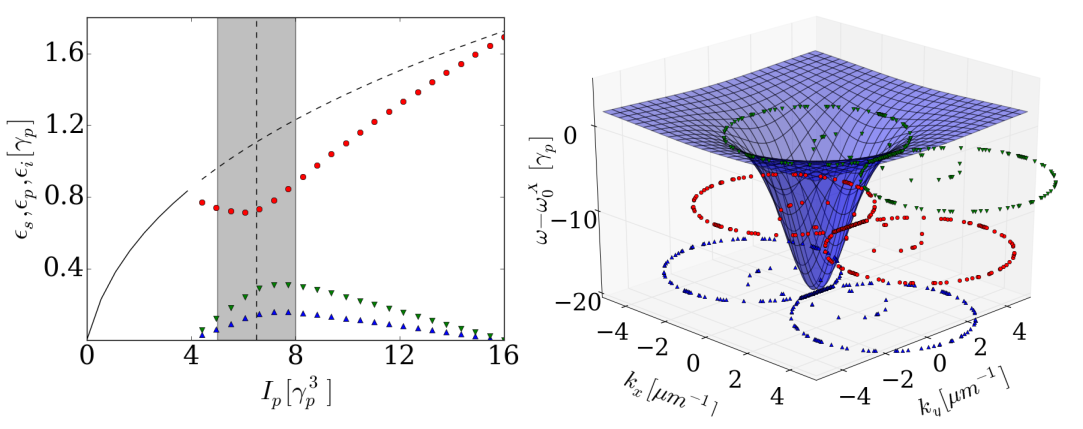}
\caption{(Color online) OPO mean-field blue-shifts and fluctuation
  Rayleigh rings in the linear response scheme for homogeneous
  pumping. Left panel: Signal $s$ ([blue] upper triangles), pump $p$
  ([red] circles), and idler $i$ ([green] lower triangles) mean-field
  energy blue-shifts $\epsilon_{n=s,p,i}$ (in units of $\gamma_p =
  \gamma_{\vect{k}_p}$) versus the rescaled pump intensity $I_p$ (in
  units of $\gamma_p^3$) in the optical limiter regime. Parameters are
  $\Omega_R=5$~meV, zero cavity-exciton detuning, $\gamma_X = \gamma_C
  = 0.12$~meV, $\omega_p - \omega_0^X = -1.25$~meV, $k_p=1.6{\mu
    m}^{-1}$, $k_s \simeq 0$, and $k_i=3.2{\mu m}^{-1}$.  The shaded
  area is stable OPO region, while the vertical dashed line
  corresponds to the pump power value chosen for plotting the right
  panel. Right panel: Blue-shifted LP dispersion~\eqref{eq:blues} with
  superimposed Rayleigh curves $\Gamma_{p,i,(u,v), \tilde{\vect{k}} +
    \vect{k}_{p, i}}$ evaluated within the linear response
  approximation (same symbols as left panel). The two rings
  corresponding to the signal state, $\Gamma_{s,(u,v),
    \tilde{\vect{k}}}$, are shrinked to zero because $k_s \simeq 0$.}
\label{fig:spect}
\end{figure}
%
\section{Model}
The dynamics of polaritons in the OPO regime, and their hydrodynamic
properties when scattering against a defect, can be described via a
classical driven-dissipative non-linear Gross-Pitaevskii equation
(GPE) for the coupled exciton and cavity fields $\psi_{X,C}
(\vect{r},t)$ ($\hbar=1$)~\cite{whittaker2005_b,Carusotto2013a}:
\begin{equation}
  i\partial_t \begin{pmatrix} \psi_X \\ \psi_C \end{pmatrix} =
  \hat{H} \begin{pmatrix} \psi_X \\ \psi_C \end{pmatrix}
  + \begin{pmatrix} 0 \\ F_p(\vect{r},t) \end{pmatrix} \; .
\label{eq:gpequ}
\end{equation}
The dispersive $X$- and $C$-fields decay at a rate $\gamma_{X,C}$ and
are coupled by the Rabi splitting $\Omega_R$, while the non-linearity
is regulated by the exciton coupling strength $g_X$:
\begin{equation}
  \hat{H} = \begin{pmatrix} \omega^{X}_{-i\nabla} - i
    \frac{\gamma_X}{2} + g_X |\psi_X|^2 & \Omega_R/2 \\ \Omega_R/2 &
    \omega^C_{-i\nabla} - i \frac{\gamma_C}{2} + V_d \end{pmatrix} \;
  .
\end{equation}
We describe the defect via a potential $V_d (\vect{r})$ acting on the
photonic component; this can either be a defect in the cavity mirror
or a localised laser field~\cite{amo09_b,amo10,zajac2012}.
In the conservative, homogeneous, and linear regime ($\gamma_{X,C}=0=
V_d (\vect{r})= g_X$), the eigenvalues of $\hat{H}$ are given by the
lower (LP) and upper polariton (UP) energies, $2
\omega_{\vect{k}}^{LP,UP} = \omega_{\vect{k}}^{C} +
\omega_{\vect{k}}^{X} \mp \sqrt{(\omega_{\vect{k}}^{C} -
  \omega_{\vect{k}}^{X})^2 + \Omega_R^2}$.
The cavity is driven by a continuous-wave laser field $F_p(\vect{r},t)
= \mathcal{F}_p(\vect{r}) e^{i (\vect{k}_p \cdot \vect{r} - \omega_p
  t)}$ into the OPO regime: Here, polaritons are continuously injected
into the pump state with frequency $\omega_p$ and momentum
$\vect{k}_p$ and, above a pump strength threshold, undergo coherent
stimulated scattering into the signal $(\omega_s, \vect{k}_s)$ and
idler $(\omega_i, \vect{k}_i)$ states. 

As a first step, it is useful to get insight into the system behaviour
in the simple case of a homogeneous pump of strength
$\mathcal{F}_p(\vect{r}) = f_p$. A numerical study of the coupled
equations~\eqref{eq:gpequ} for the more realistic case of a
finite-size top-hat pump profile $\mathcal{F}_p(\vect{r})$ will be
presented later.
To further simplify our analysis, we assume here that the UP
dispersion does not get populated by parametric scattering processes
and thus, by means of the Hopfield coefficients $2X_{\vect{k}}^2,
2C_{\vect{k}}^2 = 1 \pm (\omega_{\vect{k}}^{C} -
\omega_{\vect{k}}^{X})/\sqrt{(\omega_{\vect{k}}^{C} -
  \omega_{\vect{k}}^{X})^2 + \Omega_R^2}$, we project the
GPE~\eqref{eq:gpequ} onto the LP
component~\cite{ciuti01,wouters07:prb} $\psi_{\vect{k}}^{} =
X_{\vect{k}} \psi_{X,\vect{k}} + C_{\vect{k}} \psi_{C,\vect{k}}$,
where $\psi(\vect{r},t) = \sum_{\vect{k}} e^{i\vect{k}\cdot \vect{r}}
\psi_{\vect{k}}^{} (t)$:
\begin{multline}
  i\partial_t \psi_{\vect{k}}^{} = \left[\omega_{\vect{k}}^{LP} -
    i\Frac{\gamma_{\vect{k}}}{2}\right]\psi_{\vect{k}}^{} +
  C_{\vect{k}} \sum_{\vect{q}} C_{\vect{q}} V_d(\vect{k} - \vect{q})
  \psi_{\vect{q}}^{}\\ + \sum_{\vect{k}_1, \vect{k}_2} g_{\vect{k},
    \vect{k}_1, \vect{k}_2} \psi^*_{\vect{k}_1 + \vect{k}_2-\vect{k}}
  \psi_{\vect{k}_1}^{} \psi_{\vect{k}_2}^{} + \tilde{f}_p(t)
  \delta_{\vect{k},\vect{k}_p}\; .
\label{eq:efflp}
\end{multline}
Here, $\gamma_{\vect{k}}=\gamma_X X_{\vect{k}}^2 + \gamma_C
C_{\vect{k}}^2$ is the effective LP decay rate, the interaction
strength is given by $g_{\vect{k}, \vect{k}_1, \vect{k}_2}=g_X
X_{\vect{k}} X_{\vect{k}_1 + \vect{k}_2-\vect{k}} X_{\vect{k}_1}
X_{\vect{k}_2}$, and the pumping term by $\tilde{f}_p(t)=
C_{\vect{k}_p} f_p e^{-i\omega_p t}$.

\section{Linear response theory}
In the limit where the homogeneously pumped system is only weakly
perturbed by the external potential $V_d(\vect{r})$, we apply a linear
response analysis~\cite{astrakharchik04}: The LP field is expanded
around the mean-field terms for the three $n=1,2,3=s,p,i$ OPO
states~\cite{whittaker05},
\begin{equation}
  \psi_{\tilde{\vect{k}}}^{} = \sum_{n=1}^3 e^{-i \omega_n t}
  \left[\psi_{n}^{} \delta_{\tilde{\vect{k}},0} +
    u_{n,\tilde{\vect{k}}}^{} e^{-i\omega t} +
    v^*_{n,-\tilde{\vect{k}}} e^{i\omega t} \right]\; ,
\label{eq:expan}
\end{equation}
where $\tilde{\vect{k}} = \vect{k} - \vect{k}_n$. Eq.~\eqref{eq:efflp}
is expanded linearly in both the fluctuations terms,
$u_{n,\tilde{\vect{k}}}^{}$ and $v_{n,\tilde{\vect{k}}}^{}$, as well
as the defect potential.  At zero-th order, the three complex uniform
mean-field equations can be solved to obtain the dependence of the
signal, pump and idler energy blue-shifts, $\epsilon_n = g_X
X^2_{\vect{k}_n} |\psi_{n}^{}|^2$ on the system
parameters~\cite{wouters07:prb,SM}. A typical behaviour of
$\epsilon_n$ as a function of the rescaled pump intensity $I_p = g_X
C_{\vect{k}_p}^2 f_p^2/X_{\vect{k}_p}^2$ in the optical limiter regime
is plotted in the left panel of Fig.~\ref{fig:spect}.
At first order, one obtains six coupled equations diagonal in momentum
space~\cite{wouters06b}
\begin{equation}
  \omega \vect{w}_{\tilde{\vect{k}}} = \mathcal{L}_{\tilde{\vect{k}}}
  \vect{w}_{\tilde{\vect{k}}} + \Frac{1}{2} \vectgr{\Psi}_d\; ,
\label{eq:fluct}
\end{equation}
for the $6$-component vector $\vect{w}_{\tilde{\vect{k}}} =
(u_{n,\tilde{\vect{k}}}^{} , v_{n,\tilde{\vect{k}}}^{})^{T}$ and for
the potential part, $\vectgr{\Psi}_d = (\psi_n^{} C_{\vect{k}_n}
C_{\vect{k} + \vect{k}_n} V_d(\vect{k}),-\psi_n^* C_{\vect{k}_n}
C_{\vect{k}_n - \vect{k}} V_d(-\vect{k}))^{T}$. In ~\eqref{eq:fluct}
we have only kept the terms oscillating at the frequencies $\omega_n
\pm \omega$ and neglected the other terms in the expansion (i.e.,
$2\omega_n - \omega_m \pm \omega$), which are oscillating at
frequencies far from the LP band, and thus with negligible amplitudes.
In the particle-like and the hole-like channels, the Bogoliubov matrix
determining the spectrum of excitations can be written
as~\cite{wouters06b}
\begin{equation}
  \mathcal{L}_{\vect{k}} = \begin{pmatrix} M_{\vect{k}}^{} & Q_{\vect{k}}^{}
    \\ -Q_{-\vect{k}}^* & -M_{-\vect{k}}^* \end{pmatrix} \; ,
\end{equation}
where the 3 OPO states components are
\begin{align}
  (M_{\vect{k}}^{})_{mn} &= \left[\omega^{LP}_{\vect{k}_m + \vect{k}}
    - \omega_m - i\Frac{\gamma_{\vect{k}_m + \vect{k}}}{2}\right]
  \delta_{m,n} \\
  \nonumber & + 2\sum_{q,t=1}^3
  g_{\vect{k}_m+\vect{k},\vect{k}_n+\vect{k},\vect{k}_t} \psi_q^*
  \psi_t^{} \delta_{m+q,n+t}\\
  (Q_{\vect{k}}^{})_{mn} &= \sum_{q,t=1}^3
  g_{\vect{k}_m+\vect{k},\vect{k}_q,\vect{k}_t} \psi_q^{} \psi_t^{}
  \delta_{m+n,q+t}\; .
\end{align}
\begin{figure}[h!]
\centering
\includegraphics[width=0.85\columnwidth]{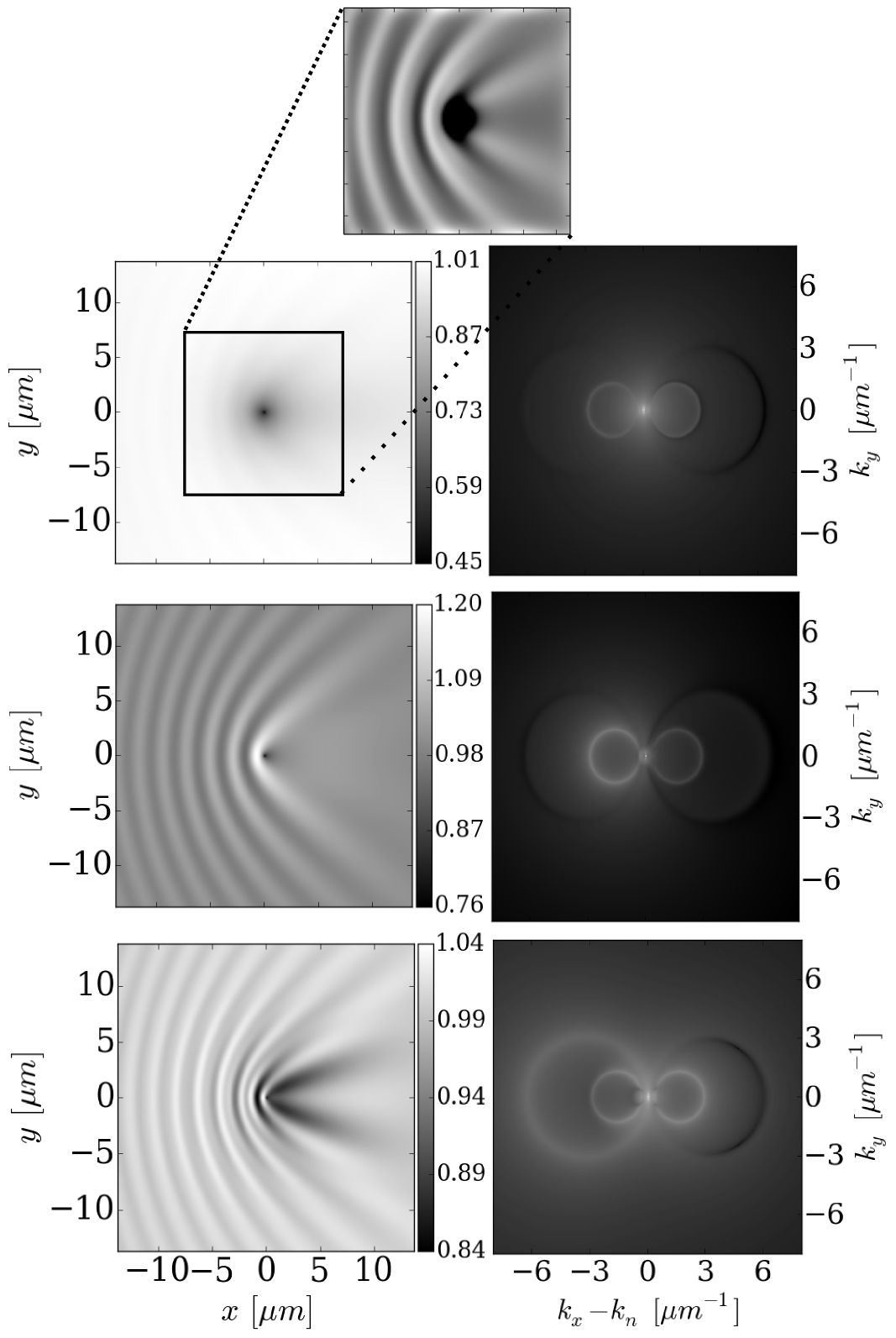}
\caption{Linear responses to a static defect of the three OPO states
  in real and momentum space. Rescaled filtered OPO emissions (signal
  [top panels], pump [middle], and idler [bottom]) in real space
  $|\psi (\vect{r},\omega_n)|^2/|\psi_n|^2$ (left panels in linear
  scale) and momentum space $|\psi_{\tilde{\vect{k}}} (\omega_n)|^2$
  (right panel in logarithmic scale) obtained within the linear
  response approximation. The parameters are the same ones of
  Fig.~\ref{fig:spect}. For the top left panel of the signal emission
  in real space, Gaussian filtering is applied to enhance the short
  wavelength modulations, which amplitudes are otherwise roughly $1\%$
  of the average signal intensity and about a factor of 10 times
  weaker than the modulation amplitudes in the pump fluid.}
\label{fig:ereal}
\end{figure}
In absence of a defect potential ($\vectgr{\Psi}_d =0$),
Eq.~\eqref{eq:fluct} is the eigenvalue equation for the spectrum of
excitations of a homogeneous OPO, i.e.,
$\det(\mathcal{L}_{\tilde{\vect{k}}} - \omega)=0$. The spectrum has 6
branches, $\omega_{n,(u,v),\tilde{\vect{k}}}$, labeled by $n=s,p,i$
and $(u,v)$. Even though these degrees of freedom are mixed together,
at large momenta, one recovers the LP dispersions shifted by the three
states energies and momenta, i.e.,:
\begin{equation}
  \lim_{\tilde{k} \gg \sqrt{2m_C \Omega_R}} \omega_{n,(u,v),\tilde{\vect{k}}} = \pm
  (\omega^{LP}_{\vect{k} - \vect{k}_n} - \omega_n)\; ,
\label{eq:large}
\end{equation}
where $+$ ($-$) corresponds to the $u$ ($v$) particle- (hole-)like
branch.
The OPO solution is stable (shaded area in Fig.~\ref{fig:spect}) as
far as $\Im \omega_{n,(u,v),\tilde{\vect{k}}} < 0$.

The shape of the patterns, or Cherenkov-like waves, resulting from the
elastic scattering of the OPO 3-fluids against the static ($\omega=0$)
defect can be determined starting from the spectrum, and in particular
evaluating the closed curves $\Gamma_{n,(u,v), \tilde{\vect{k}}}$ in
$\vect{k}$-space, or ``Rayleigh rings''~\cite{Rousseaux2011} defined
by condition $\Re \omega_{n,(u,v), \tilde{\vect{k}}} =
0$~\footnote{Even if they do not appear to be relevant here, note that
  the presence of a non-vanishing imaginary part of the excitation
  spectrum $Im\omega_{n,(u,v),\tilde{\vect{k}}}\neq 0$ introduces some
  complications: Even in the absence of any Rayleigh ring, the drag
  force can be non-vanishing and the standard Landau criterion may
  fail identifying a critical velocity~\cite{wouters2010}.}.
The modulations propagate with a direction
$\hat{\eta}_{n,(u,v),\tilde{\vect{k}}}$ orthogonal to each curve
$\Gamma_{n,(u,v), \tilde{\vect{k}}}$, a pattern wavelength given by
the corresponding $|\tilde{\vect{k}}|$, and a group velocity
$\vect{v}_{n,(u,v),\tilde{\vect{k}}}^{(g)}=\nabla_{\tilde{\vect{k}}}
\Re \omega_{n,(u,v), \tilde{\vect{k}}}$, where
$\xi_{n,(u,v),\tilde{\vect{k}}} =
|\vect{v}_{n,(u,v),\tilde{\vect{k}}}^{(g)} / \Im \omega_{n,(u,v),
  \tilde{\vect{k}}}|$ determines the distance, at any given direction
$\hat{\eta}_{n,(u,v),\tilde{\vect{k}}}$, over which the perturbation
extends away from the defect. For a single fluid under a coherent
pump, the qualitative shape of the modulation pattern generated in the
fluid by the defect is mostly determined by the excitation
spectrum~\cite{carusotto06_prl,carusotto04}.

For OPO, the spectrum of excitation on top of each of the three,
$n=1,2,3$, states (see~\cite{SM}) generates six identical Rayleigh
rings $\Gamma_{n,(u,v), \tilde{\vect{k}}}$ for the 3 states.
The Rayleigh rings for the OPO conditions specified in
Fig.~\ref{fig:spect} are clearly visible in the right panels of
Fig.~\ref{fig:ereal}, where we plot the $\vect{k}$-space
photoluminescence filtered at each state energy, i.e.,
$|\psi_{\tilde{\vect{k}}} (\omega_n)|^2= |\psi_{n}^{}
\delta_{\tilde{\vect{k}},0} + u_{n,\tilde{\vect{k}}}^{} +
v^*_{n,-\tilde{\vect{k}}}|^2$.
We have here chosen a $\delta$-like defect potential, $V_d(\vect{k}) =
g_d$, but we have however checked that our results do not depend on
its exact shape~\cite{SM}.
For the OPO conditions considered here, the signal momentum is at $k_s
\simeq 0$, and thus only four of the six rings are present. The same
rings are also plotted in the right panel of Fig.~\ref{fig:spect},
shifted at each of the three OPO state momentum $\vect{k}_n$,
$\Gamma_{n,(u,v), \tilde{\vect{k}}+\vect{k}_n}$ and energies
$\omega_n$.
It is important to note that, even though the three OPO states have
locked responses because the three states display the same spectrum of
excitations, only one of the rings $\Gamma_{n,(u,v),
  \tilde{\vect{k}}+\vect{k}_n}$ is the most resonant at $\omega_n$
with the interaction blue-shifted lower polariton dispersion,
\begin{equation}
  \bar{\omega}_{\vect{k}}^{LP} = \omega_{\vect{k}}^{LP} + 2 
  X_{\vect{k}}^2 \sum_{n=1}^3  \epsilon_n\; ,
\label{eq:blues}
\end{equation}
where $\epsilon_n = g_X X_{\vect{k}_n}^2 |\psi_n^{}|^2$ are the
mean-field energy blue-shifts (measured in Fig.~\ref{fig:spect} in
units of $\gamma_p = \gamma_{\vect{k}_p}$).  This implies that the
most visible modulation for each fluid should be the most resonant
one, with superimposed weaker modulations coming from the other two
state rings.

In the specific case of Fig.~\ref{fig:spect}, the signal is at $k_s
\simeq 0$ and thus produces no rings in momentum space. The other four
rings are very far from being resonant with the blue-shifted LP
dispersion~\eqref{eq:blues} at $\omega_s$, and thus the signal
displays only an extremely weak modulation coming from the next
closer ring, which is the one associated with the pump state,
$\Gamma_{p,u, \tilde{\vect{k}}+\vect{k}_s}$.
We estimate that the signal modulation amplitudes are roughly $1\%$ of
the average signal intensity and about a factor of 10 times weaker
than the modulation amplitudes in the pump fluid.
In order to show that the signal has weak modulations coming from
the pump, we apply a Gaussian filter to the real space images (see
inset of the left top panel of Fig.~\ref{fig:ereal}).
As explained in more details in~\cite{SM}, Gaussian filtering consists
of subtracting from the original data a copy which has been
convoluted with a Gaussian kernel, thus getting rid of the
long-wavelength modulations.
This procedure reveals that indeed the pump imprints its modulation
also into the signal, even though these are extremely weak, thus
leaving the signal basically insensitive to the presence of the
defect.

Pump and idler states are each mostly resonant with their own
rings, i.e., $\Gamma_{p,u, \tilde{\vect{k}}+\vect{k}_p}$ at $\omega_p$
and $\Gamma_{i,u, \tilde{\vect{k}}+\vect{k}_i}$ at $\omega_i$,
respectively. Thus one should then observe two superimposed
modulations in both pump and idler filtered emissions, the stronger
one for each being the most resonant one.
However, the modulations associated to the idler only propagate very
close to the defect, at an average distance
$\overline{\xi_{i,u,\tilde{\vect{k}}}} \sim 1.7~\mu$m before getting
damped, and thus are not clearly visible. For the OPO conditions
considered, this is due to the small idler group velocity
$\vect{v}_{i,u,\tilde{\vect{k}}}^{(g)}$, as the dispersion is almost
excitonic at the idler energy.

We can conclude that, for the typical OPO condition with a signal at
$k_s \simeq 0$, considered in Figs.~\ref{fig:spect}
and~\ref{fig:ereal}, the signal fluid does not show modulations and
the extremely weak scattering inherited from the pump state can be
appreciated only after a Gaussian filtering procedure of the image. In
contrast, the idler has a locked response to the one of the pump
state.
Note that, for the conditions shown in Fig.~\ref{fig:spect}, as well
as the other cases considered in Ref.~\cite{SM}, the subsonic to
supersonic crossover of the pump-only state~\cite{amo09_b} happens
well above the region of stability of OPO. Thus it is not possible to
study a case where the pump is already subsonic and at the same time
promotes stimulated scattering.
\begin{figure}[h!]
\centering \includegraphics[width=0.8\columnwidth]{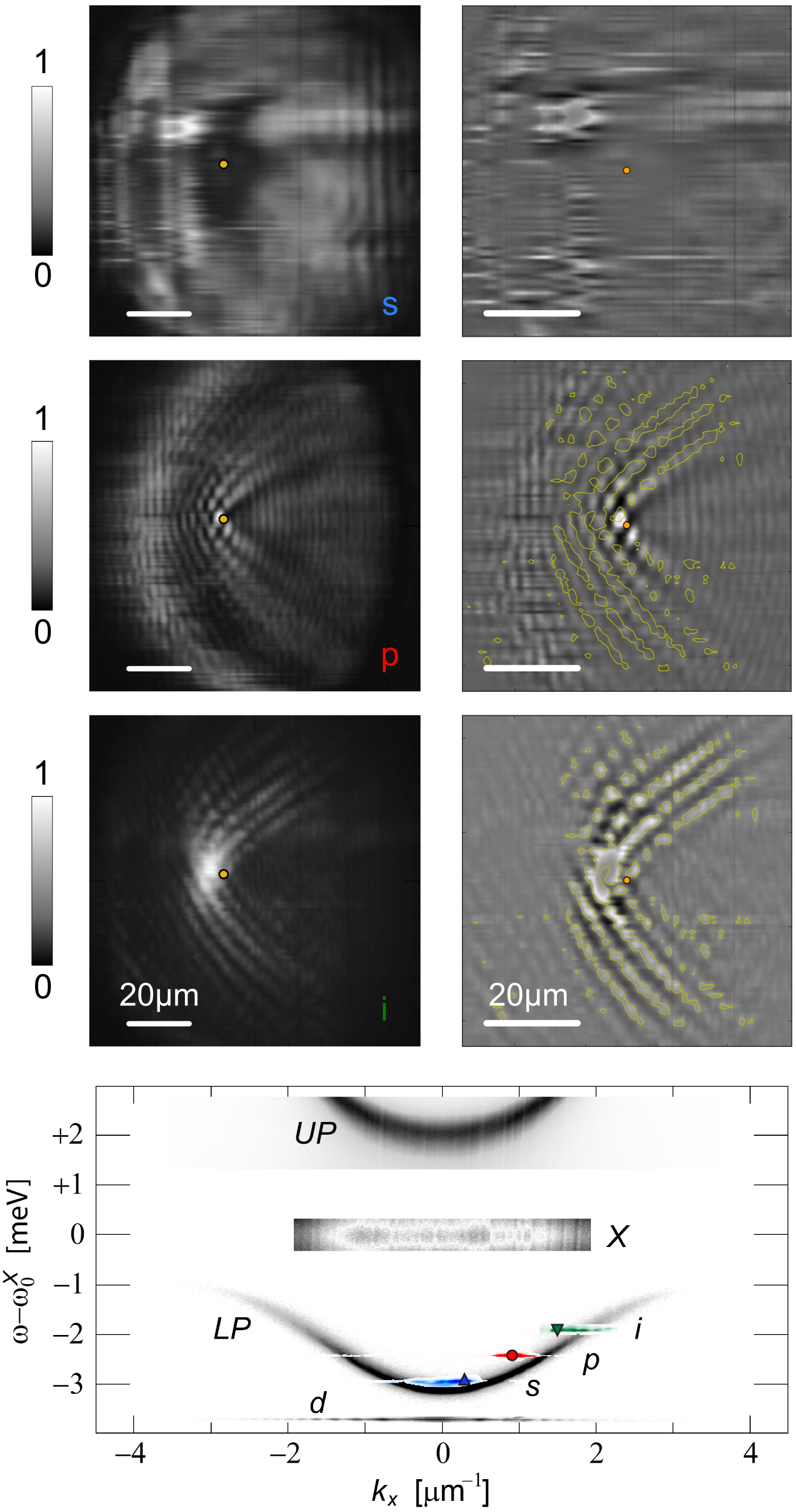}
\caption{(Color online) Experimental OPO spectrum and filtered
  emissions of signal, pump and idler in presence of a structural
  defect. The six panels show the filtered emission profiles in real
  space of signal (top), pump (middle) and idler (bottom) $I_{s,
    p,i}(\vect{r})$. A Gaussian filtering to enhance the short
  wavelength modulations is applied in the right column panels. Here,
  the extracted wave-crests from the idler emission (yellow contours
  in the bottom panel) are also superimposed to the pump profile
  (middle) by applying a $\pi$-phase shift. The (orange) dot indicates
  the position of the defect. The lower panel shows the experimental
  OPO spectrum. Energy and momentum of the three OPO states are
  labeled with a [blue] upper triangle (signal), a [red] circle
  (pump), and a [green] lower triangle (idler), while the localised
  state, clearly visible just below the bottom of the LP dispersion,
  is indicated with the symbol $d$. The bare LP dispersion is
  extracted from an off-resonant low pump power measurement, as well
  as the emission of the exciton reservoir (X) and the one of the UP
  dispersion (each in a different scale).}
\label{fig:exper}
\end{figure}
%
\section{Experiments}
We now turn to the experimental analysis. We use a continuous-wave
laser to drive a high quality ($Q=14000$) GaAs microcavity sample into
the OPO regime --- details on the sample can be found in a previous
publication~\cite{ballarini2013,dominici14}.
The polariton dispersion is characterised by a Rabi splitting
$\Omega_R=5.4$~meV, the exciton energy $\omega_0^{X}=1485.26$~meV and
we choose a sample region where the cavity-exciton detuning is
slightly negative, $-1$~meV. We pump at $k_p=0.89~\mu$m$^{-1}$ and
$\omega_p - \omega_0^{X}=-2.43$~meV, and, at pump powers $1.5$-times
above threshold, we obtain an OPO with signal at small wavevector
$k_s=0.21~\mu$m$^{-1}$ and $\omega_s - \omega_0^{X}=-2.95$~meV, and
idler at $k_i=1.57~\mu$m$^{-1}$ and $\omega_i -
\omega_0^{X}=-1.91$~meV.
The defect we use in the sample is a localized inhomogeneity naturally
present in the cavity mirror. Note that the exact location of the
defect can be extracted from the emission spectrum and is indicated
with a dot (orange) symbol in the profiles of Fig.~\ref{fig:exper}.

In order to filter the emission at the three states energies,
$I_{s,p,i}(\vect{r}=x,y)$, and obtain 2D spatial maps for the OPO
three states, we use a spectrometer and, at a fixed position $x_0$,
obtain the intensity emission as a function of energy and position,
$I(\epsilon,x_0,y)$. By changing $x_0$ we build the full emission
spectrum as a function of energy and 2D position,
$I(\epsilon,\vect{r})$. The filtered emission for each OPO state is
obtained from the integrals $I_{n=s,p,i}(\vect{r}) =
\int_{\omega_n-\sigma}^{\omega_n+\sigma} d\epsilon I(\epsilon,
\vect{r})$, with $\sigma=0.08$~meV. The results are shown in
Fig.~\ref{fig:exper} for respectively the signal (top panel), pump
(middle) and the idler (bottom) profiles.
The signal profile shows no appreciable modulations around the defect
locations, nor these could be observed after applying a Gaussian
filtering procedure of the image.
In contrast, in agreement with the theoretical results, both filtered
profiles of pump and idler show the same Cherenkov-like pattern. We
extract the wave-crests from the idler profile ([yellow] contours in
the bottom panel) and superimpose them to the pump profile (middle
panel) with an added $\pi$-phase-shift, revealing that the only
modulations visible in the idler state are the ones coming from the
pump state.

\section{Numerical analysis}
The agreement between the results obtained experimentally and within
the linear response approximation is additionally confirmed by an
exact full numerical analysis of the coupled
equations~\eqref{eq:gpequ} for a finite size pump via a
5$^{\text{th}}$-order adaptive-step Runge-Kutta algorithm.
Details are given in~\cite{SM}.
The pumping conditions are very similar to those previously considered
in the linear response approximation of Figs.~\ref{fig:spect}
and~\ref{fig:ereal}, while the pump profile $\mathcal{F}_p(\vect{r})$
is now finite-size top-hat. In particular, we consider the case of
zero cavity-exciton detuning, $k_p=1.6$~$\mu$m$^{-1}$,
$\omega_p-\omega_X^0 = -0.44$~meV and the pump power strength is fixed
just above threshold, so that to produce a stable steady state OPO
with, in absence of the defect, signal at small wavevector
$k_s=-0.2$~$\mu$m$^{-1}$ and idler at $k_i=3.4$~$\mu$m$^{-1}$.

\begin{figure}[h!]
\centering
\includegraphics[width=0.85\columnwidth]{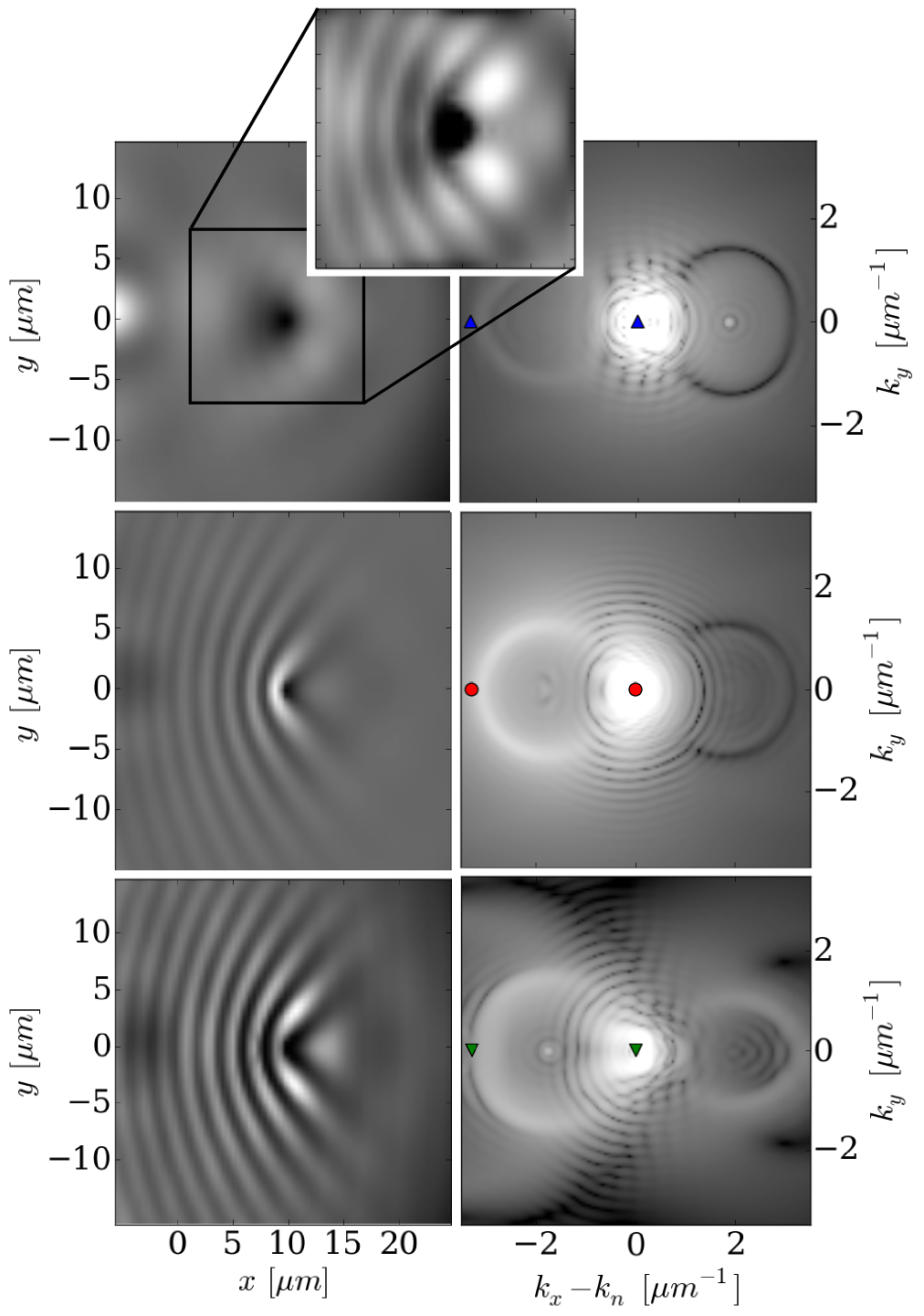}
\caption{(Color online) Full numerical responses to a static defect of
  the three OPO states in real and momentum space. Filtered OPO
  emissions (signal [top panels], pump [middle], and idler [bottom])
  in real space $|\psi_C (\vect{r},\omega_n)|^2$ (left panels in
  linear scale) and momentum space $|\psi_C
  (\tilde{\vect{k}},\omega_n)|^2$ (right panel in logarithmic scale)
  obtained by a full numerical evaluation of~\eqref{eq:gpequ}. For the
  top left panel of the signal space emission, Gaussian filtering is
  applied to enhance the short wavelength modulations of this state,
  revealing that the modulations corresponding to the pump state are
  also imprinted (though weakly) into the signal. The symbols indicate
  the pump ring diameter extracted from fitting the upstream
  modulations and resulting in a density wave wave-vector coinciding
  with the one of the pump $k_p=1.6$~$\mu$m$^{-1}$.}
\label{fig:numer}
\end{figure}

When adding a localised defect potential, the steady state OPO
develops Rayleigh rings in momentum space, yet, as shown in~\cite{SM},
the spectrum continues to be $\delta$-like in energy, allowing to
easily filter in energy the emission of the three OPO states.
Results are shown in Fig.~\ref{fig:numer}, where real-space emissions
$|\psi_C (\vect{r},\omega_n)|^2$ are plotted in the left panels, while
the ones in momentum space $|\psi_C (\tilde{\vect{k}},\omega_n)|^2$ in
the right panels. We observe a very similar phenomenology to that one
obtained in the linear approximation shown in
Fig.~\ref{fig:ereal}. The signal now is at slightly negative values of
momenta $k_s=-0.2$~$\mu$m$^{-1}$, thus implying a very small Rayleigh
ring associated with this state. Thus we observe that only the
modulations associated with the pump are the ones that are weakly
imprinted in the signal state and that can be observed by means of a
Gaussian filtering (inset of top-left panel). We have fitted the
upstream wave-crests and obtained the same modulation wave-vector as
the pump one ([blue] upper triangles). Similarly to the linear
response case, we also find here that the most visible perturbation in
the emission filtered at the idler energy is the one due to the pump
Rayleigh ring. As before, the modulations due to the idler Rayleigh
ring cannot propagate far from the defect because of the small group
velocity associated with this state.

%
\section{Conclusions}
To conclude, we have reported a joint theoretical and experimental
study of the superfluid properties of a non-equilibrium condensate of
polaritons in the so-called optical parametric oscillator
configuration by studying the scattering against a static defect.
We have found that while the signal is basically free from
modulations, pump and idler lock to the same response. We have
highlighted the role of the coupling between the OPO components by
non-linear and parametric processes. These are responsible for the
transfer of the spatial modulations from one component to the
other. This process is most visible in the clear spatial modulation
pattern that is induced by the non-superfluid pump onto the idler,
while the same modulations are only extremely weakly transferred into
the signal, because of its low characteristic wavevector, so much that
experimentally cannot be resolved.
The main features of the real- and momentum-space emission patterns
are understood in terms of Rayleigh scattering rings for each
component and a characteristic propagation length from the defect; the
rings are then transferred to the other components by nonlinear and
parametric processes.
Our theoretical and experimental results stress the complexities and
richness involved when looking for superfluid behaviours in
non-equilibrium multicomponent condensates such as the ones obtained
in the optical parametric oscillation regime.

\acknowledgments We are grateful to M. Wouters, C. Tejedor and M. De
Giorgi for stimulating discussions.
Financial support from the ERC POLAFLOW (grant n. 308136) is
acknowledged.
FMM acknowledges financial support from the Ministerio de Econom\'ia y
Competitividad (MINECO, contract No. MAT2011-22997), the Comunidad
Autonoma de Madrid (CAM, contract No. S-2009/ESP-1503), and the
European Science Foundation (ESF) program Intelbiomat.
IC acknowledges financial support by the ERC through the QGBE grant
and by the Autonomous Province of Trento, partly through the ``On
silicon chip quantum optics for quantum computing and secure
communications'' (``SiQuro'').
MHS acknowledges support from EPSRC (grants EP/I028900/2 and
EP/K003623/2).

\newpage

\pagenumbering{gobble}

\setcounter{secnumdepth}{2}

\renewcommand{\figurename}{\textsc{S.~Fig.}}

\setcounter{equation}{0}
\setcounter{figure}{0}

\onecolumngrid
\begin{center}
\large
\textbf{Supplemental Material for ``On multicomponent polariton
  superfluidity in the optical parametric oscillator regime''}  
\end{center}
\vspace{3ex}
\twocolumngrid

In the manuscript, we carry on a theoretical and experimental analysis
of the response of microcavity polaritons in the optical parametric
oscillator (OPO) regime to a static defect.
For the theoretical calculations we follow two independent approaches:
In the first approach, we exactly numerically solve the dynamics of
the two coupled Gross-Pitaevskii equations (GPEs) for the exciton and
cavity fields for the realistic case of a finite-size top-hat pump
profile $\mathcal{F}_p(\vect{r})$. In the second approach, we apply a
perturbative linear response approximation for the lower polariton
(LP) state which leads to semi-analytical results in the limiting case
of a spatially homogeneous pump profile. Both methods have been
already successfully used in the literature to explore several
properties of the OPO operation.

While all fundamental formulae and information has been given in the
main text, in this Supplementary Material (SM) we present some
additional details on both approaches that might be of interest to the
specialized reader.  In addition to that, we make use of the linear
response approach to examine some regimes that are hardly accessed
experimentally or within a full numerical approach, which however
allow to put the conclusions of our work into a broader perspective.

\section{Full numerics}
\label{sec:detun}
The classical driven-dissipative non-linear Gross-Pitaevskii equation
(GPE) for the coupled exciton and cavity fields $\psi_{X,C}
(\vect{r},t)$ ($\hbar=1$)
\begin{equation}
  i\partial_t \begin{pmatrix} \psi_X \\ \psi_C \end{pmatrix} =
  \hat{H} \begin{pmatrix} \psi_X \\ \psi_C \end{pmatrix}
  + \begin{pmatrix} 0 \\ F_p(\vect{r},t) \end{pmatrix}\; ,
\label{eq:numer}
\end{equation}
where
\begin{equation}
  \hat{H} = \begin{pmatrix} \omega^{X}_{-i\nabla} - i
    \frac{\gamma_X}{2} + g_X |\psi_X|^2 & \Omega_R/2 \\ \Omega_R/2 &
    \omega^C_{-i\nabla} - i \frac{\gamma_C}{2} + V_d \end{pmatrix} \;
  ,
\end{equation}
is solved numerically on a 2D grid of $N\times N=2^8\times 2^8$ points
and a separation of $0.47$~$\mu$m (i.e., in a box $L\times L =
121$~$\mu$m$\times 121$~$\mu$m) by using a 5$^{\text{th}}$-order
adaptive-step Runge-Kutta algorithm. Convergence has been checked both
with respect the resolution in space $L/N$ as well as in momentum
$\pi/L$, without~\cite{marchetti10,marchetti_review} as well as in
presence of the defect.
The same approach has been already used in previous publications from
some of the authors (for a review, see
Refs.~\cite{marchetti10,marchetti_review}).
As for the system parameters, we have considered a LP dispersion at
zero photon-exciton detuning, $\omega^C_0 = \omega^X_0$, a
dispersionless excitonic spectrum, $\omega^X_{\vect{k}} = \omega^X_0$
and a quadratic dispersion for photons $\omega^C_{\vect{k}} =
\omega^C_0 + k^2/2m_C$, with the photon mass $m_C=2.3 \times 10^{-5}
m_e$, where $m_e$ is the bare electron mass. The LP dispersion, $2
\omega_{\vect{k}}^{LP} = \omega_{\vect{k}}^{C} + \omega_{\vect{k}}^{X}
- \sqrt{(\omega_{\vect{k}}^{C} - \omega_{\vect{k}}^{X})^2 +
  \Omega_R^2}$ is characterised by a Rabi splitting $\Omega_R =
4.4$~meV. Further, the exciton and cavity decay rates are fixed to
$\gamma_X=\gamma_C=0.53$~meV.
For the defect we choose a $\delta$-like defect potential
\begin{equation}
  V_d(\vect{r}) = g_d \delta(\vect{r} - \vect{r}_0)\; ,
\end{equation}
where its location $\vect{r}_0$ is fixed at one of the $N \times N$
points of the grid.
Note that in a finite-size OPO, local currents lead to inhomogeneous
OPO profiles in within the pump spot, despite the external pump having
a top-hat profile with a completely flat inner
region~\cite{marchetti10,marchetti_review} --- as shown later, this
can be observed in the filtered OPO profiles evaluated in absence of a
defect shown as dashed lines in S.~Fig.~\ref{fig:rafull}.
We have thus chosen the defect location so that it lies in the
smoothest and most homogeneous part of the OPO profiles.

Also we have checked that our results do not qualitatively depend on
the strength $g_d$ (nor on the sign) of the defect potential, as far
as this does not exceed a critical value above which it destabilises
the OPO steady-state regime.
The pump, $F_p(\vect{r},t) = \mathcal{F}_p(\vect{r}) e^{i (\vect{k}_p
  \cdot \vect{r} - \omega_p t)}$, has a smoothen and rotationally
symmetric top-hat profile, $\mathcal{F}_p(\vect{r}) = \mathcal{F}_p(r)
= \frac{f_p}{2}[\tanh(\frac{r+\sigma_p}{r_0}) -
  \tanh(\frac{r-\sigma_p}{r_0})]$ with strength $f_p = 1.23
f_p^\text{th} = 0.053$~meV/$\mu$m and parameters $r_0 = 8.68~\mu$m,
$\sigma_p = 34.72~\mu$m.
We pump at $k_p=1.6$~$\mu$m$^{-1}$ in the $x$-direction, $\vect{k}_p =
(k_p,0)$, and at $\omega_p-\omega_X^0=-0.44$~meV, i.e., roughly
$0.5$~meV above the bare LP dispersion. By increasing the pump
strength $f_p$, we find the threshold $f_p^{\text{th}}$ above which
OPO switches on, leading to two conjugate signal and idler states.
We then fix the pump strength just above this threshold ($f_p=1.23
f_p^{\text{th}}$), where we find a steady state OPO solution which is
stable (see Ref.~\cite{marchetti_review} for further details). In
absence of the defect, this condition corresponds to a signal state at
$k_s=-0.2$~$\mu$m$^{-1}$ and $\omega_s-\omega_0^X = -1.64$~meV and an
idler at $k_i=3.4$~$\mu$m$^{-1}$ and $\omega_i-\omega_0^X =
0.76$~meV. 
It is interesting to note that already very close to the
lower pump power threshold for OPO, the selected signal momentum is
very close to zero. This contrasts with what one obtains in the linear
approximation scheme, where instead just above the lower OPO threshold
there exists a broad interval of permitted values for $k_s$ (and thus
$k_i$) --- we will discuss further this ``selection problem'' for
parametric scattering later in the section ``Linear response''.

\begin{figure}[h!]
\includegraphics[width=1.0\columnwidth]{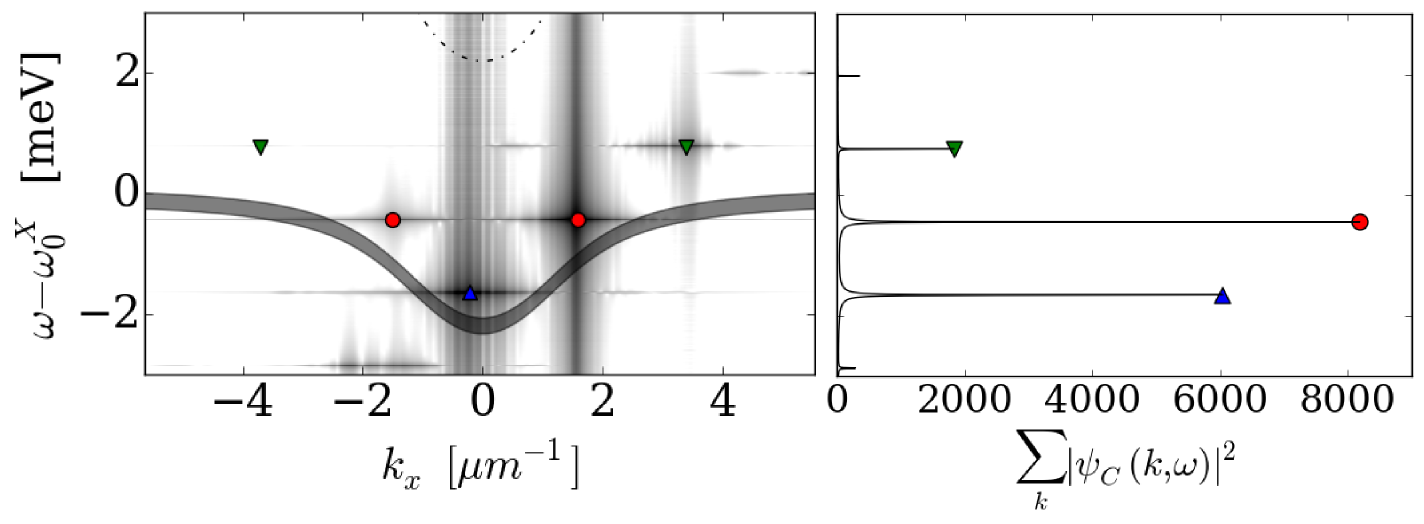}
\caption{(Color online) OPO spectrum obtained by full numerics. Left
  panel: Photonic component of the OPO spectrum in presence of a
  point-like defect, $|\psi_C(k_x,0,\omega)|^2$ (logarithmic scale),
  as a function of the rescaled energy $\omega - \omega_0^X$ versus
  the $x$-component of momentum $k_x$ (cut at $k_y=0$) for a top-hat
  pump (see text for the space profile and parameter values), with
  intensity $f_p=1.23 f_p^{\text{th}}$ above the OPO threshold, pump
  wave-vector $k_p=1.6$~$\mu$m$^{-1}$ in the $x$-direction and
  $\omega_p-\omega_X^0=-0.44$~meV. The symbols indicate the signal
  ([blue] upper triangle), pump ([red] circle), and idler ([green]
  lower triangle) energies, as well as the two momenta $k_x$ on each
  state Rayleigh ring at $k_y=0$. Note that the logarithmic scale
  results in a fictitious broadening in energy of the spectrum, which
  is in reality $\delta$-like (see right panel). The bare LP
  dispersion, including its broadening due to finite lifetime, is
  plotted as a shaded grey region, while the bare UP dispersion as a
  (black) dot-dashed line. Right panel: Momentum integrated spectrum,
  $\sum_{\vect{k}} |\psi_C(\vect{k},\omega)|^2$ (linear scale) as a
  function of the rescaled energy $\omega - \omega_0^X$, where it can
  be clearly appreciated that the emission is $\delta$-like in
  energy.}
\label{fig:spectGP}
\end{figure}
\begin{figure}[h!]
\includegraphics[width=0.8\columnwidth]{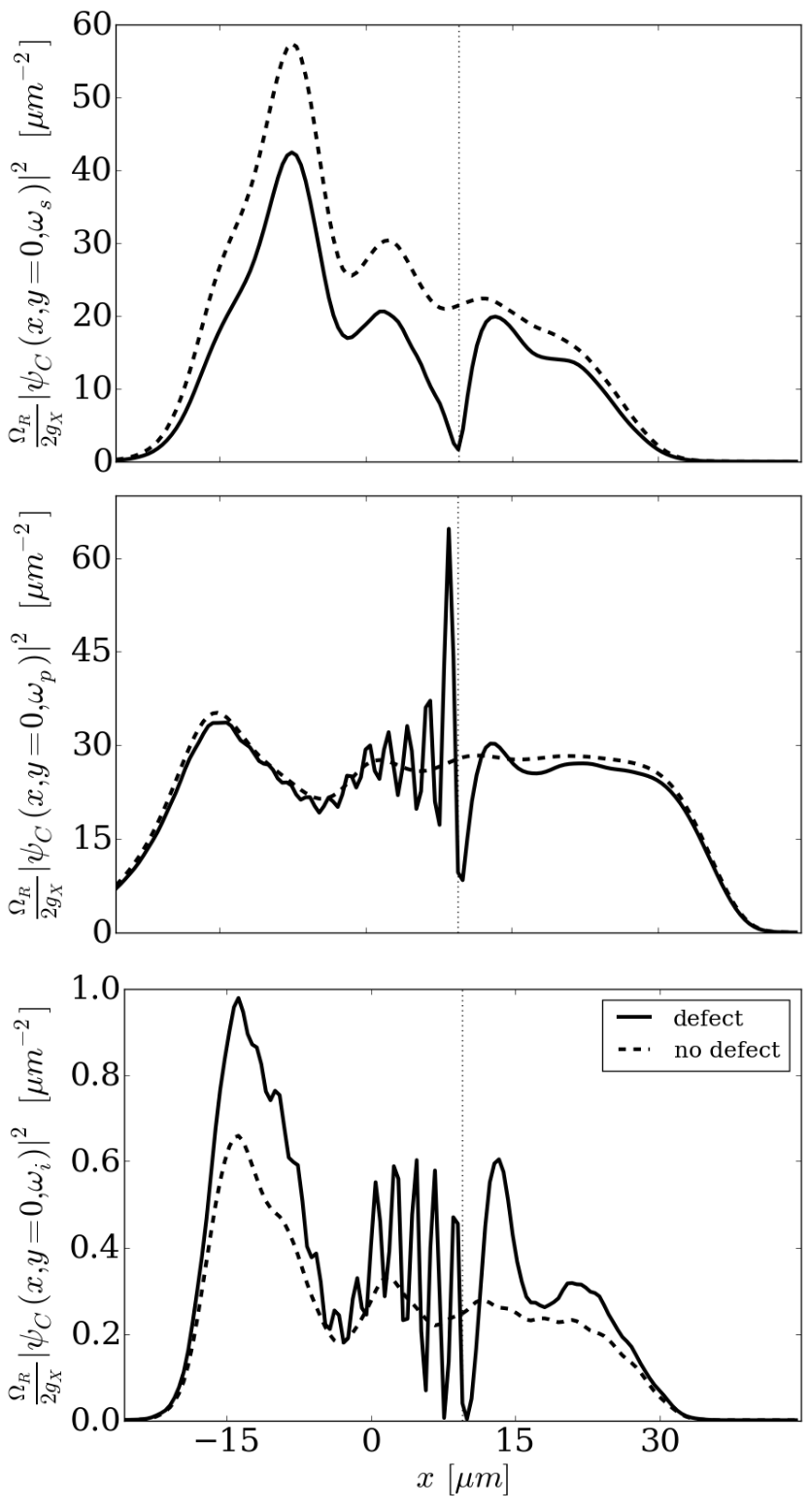}
\caption{Real space signal, pump and idler OPO one-dimensional
  filtered profiles derived from finite size numerics with and without
  a defect. Rescaled OPO filtered emissions along the $y=0$ direction,
  $|\psi_C(x,y=0,\omega_n)|^2 \frac{\Omega_R}{2g_X}$, obtained by
  numerically solving the GPE Eq.~(1) of the main manuscript. While
  the dashed lines represent the filtered emissions of signal (top
  panel), pump (middle) and idler (bottom) for a top-hat pump without
  a defect, the solid lines are the same OPO conditions but now for a
  defect positioned at $(x_d, y_d) = (9.5, -0.5)$~$\mu$m corresponding
  to the vertical dotted lines. The system parameters are the same
  ones as those of Fig.~4 of the main manuscript.}
\label{fig:rafull}
\end{figure}
We evaluate the time dependent full numerical solution
of~\eqref{eq:numer} $\psi_{X,C} (\vect{r},t)$, until a steady state
regime is reached. Here, both its Fourier transform to momentum
$\vect{k}$ and energies $\omega$ can be evaluated numerically.
We plot on the left panel of S.~Fig.~\ref{fig:spectGP} a cut at $k_y=0$
of the photonic component of the OPO spectrum in presence of a
point-like defect, $|\psi_C(k_x,0,\omega)|^2$, as a function of the
rescaled energy $\omega - \omega_0^X$ versus the $x$-component of
momentum $k_x$ (cut at $k_y=0$). In the right panel we plot instead
the corresponding momentum integrated spectrum, $\sum_{\vect{k}}
|\psi_C(\vect{k},\omega)|^2$.
Here, we can clearly see that the presence of the defect does not
modify the fact that the OPO emission for the OPO signal ([blue] upper
triangle), pump ([red] circle), and idler ([green] lower triangle)
states has a completely flat dispersion in energy, thus indicating
that a stable steady state OPO solution has been reached. Note that in
the spectrum map of the left panel of Fig.~\ref{fig:spect}, the
logarithmic scale results in a fictitious broadening in
energy. However, from the integrated spectrum plotted in linear scale
in the right panel of Fig.~\ref{fig:spect} one can clearly appreciate
that this emission is $\delta$-like, exactly as it happens for the
homogeneous OPO case~\cite{marchetti_review}.
Thus the effect of the defect is to induce only elastic (i.e., at the
same energy) scattering; now the three OPO states emit each on its own
Rayleigh ring (given each by the symbols on the left panel of
S.~Fig.~\ref{fig:spectGP} which represent the rings at a cut for
$k_y=0$). This makes it rather difficult to extract the separated
signal, pump and idler profiles by filtering in momentum, as done
previously for the homogeneous case, but it still allows to filter
those profiles very efficiently in energy. In fact, because the
emission is $\delta$-like, it is enough to fix a single value of the
energy $\omega$ to the one of the three states $\omega_{n=s,p,i}$,
thus extracting the filtered profiles either in real space
$|\psi_C(\vect{r},\omega_n)|^2$ or in momentum space
$|\psi_C(\vect{k},\omega_n)|^2$ --- we have however checked that
integrating in a narrow energy window around $\omega_n$ does not
quantitatively change the results.

The results of the above described filtering are shown in Fig.~4 of
the manuscript.
In S.~Fig.~\ref{fig:rafull} we instead plot the corresponding
one-dimensional profiles in the $y=0$ direction both in presence
(solid line) and without (dashed line) a defect. Here, we can observe
that, even if the pump has a top-hat flat profile, as also commented
previously, in absence of a defect, the finite-size OPO is
characterised by inhomogeneous profiles of signal, pump and idler
because of localised currents. Further, we observe that the presence
of a defect induces strong modulations in pump and idler.
In order to reveal that the pump also imprints its modulation into the
signal, even though these are extremely weak (and hardly visible in
both the top main panel of Fig.~4 of the manuscript and the top panel
of S.~Fig.~\ref{fig:rafull}), we apply a Gaussian filtering to the
signal images, which result is shown in the inset of the top-left
panel of Fig.~4 in the manuscript.
This consists of the following procedure. The original data for the
real space profile $\psi(\vect{r})$ are convoluted with a Gaussian
kernel $K(\vect{r} - \vect{r}')$, obtaining a new profile,
$\tilde{\psi}(\vect{r}') = \int d\vect{r} \psi (\vect{r}) K(\vect{r} -
\vect{r}')$, where short wavelengths features are smoothen out. The
convoluted image $\tilde{\psi}(\vect{r}')$ is then subtracted from the
original data, giving $\psi(\vect{r}) - \tilde{\psi}(\vect{r})$, and
effectively filtering out all long wavelength details.
The same procedure of Gaussian filtering is also applied to the signal
emission profile obtained within the linear response analysis and
shown in the inset of the top left panel of Fig.~2 of the manuscript.

\begin{figure}[h!]
\includegraphics[width=0.9\columnwidth]{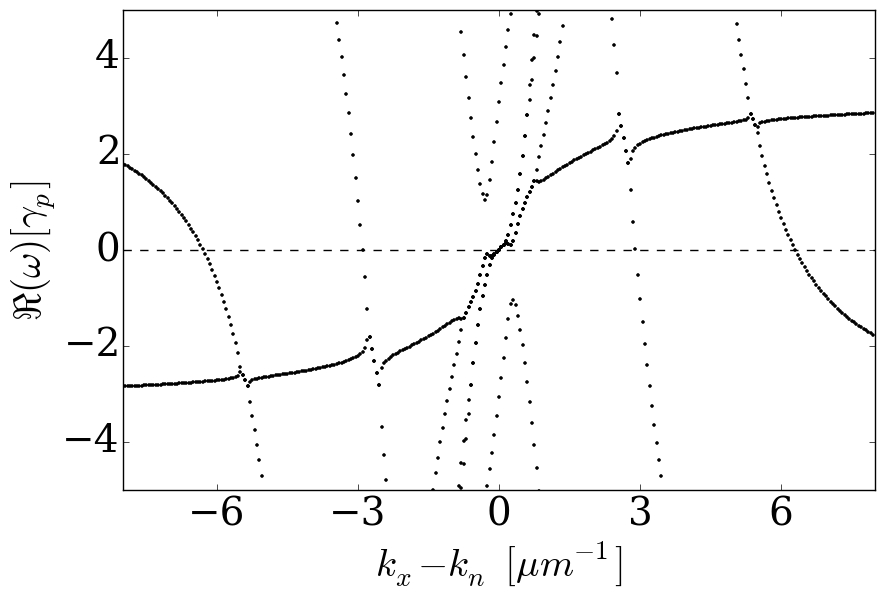}
\caption{Spectrum of collective excitations. Cut at $k_y=0$ of the
  real part of the quasiparticle energy dispersion $\Re
  \omega_{n,(u,v),\vect{k}-\vect{k}_n}$ plotted versus $k_x -
  k_n$. The spectrum is evaluated within the linear approximation
  scheme and the parameters are the same ones used for Fig.~2 of the
  main manuscript.}
\label{fig:bogol}
\end{figure}
\begin{figure}[h!]
\centering
\includegraphics[width=1.0\columnwidth]{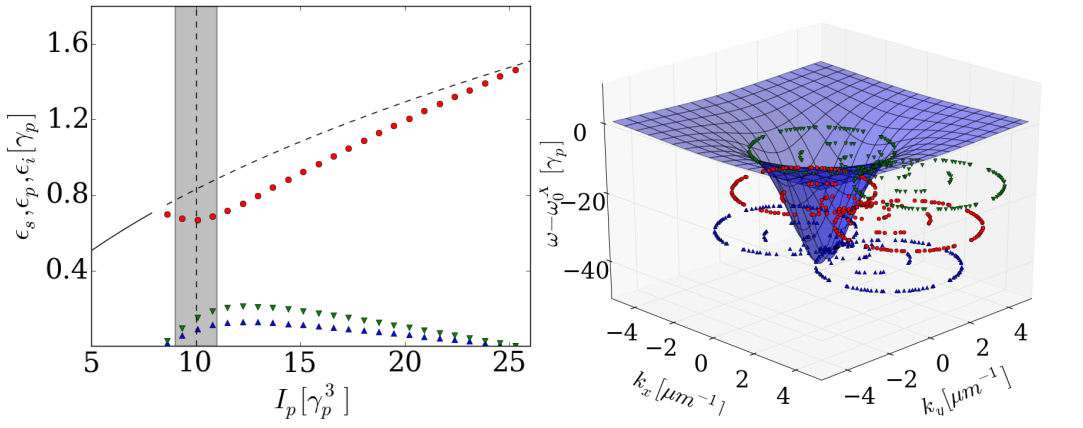}
\includegraphics[width=0.85\columnwidth]{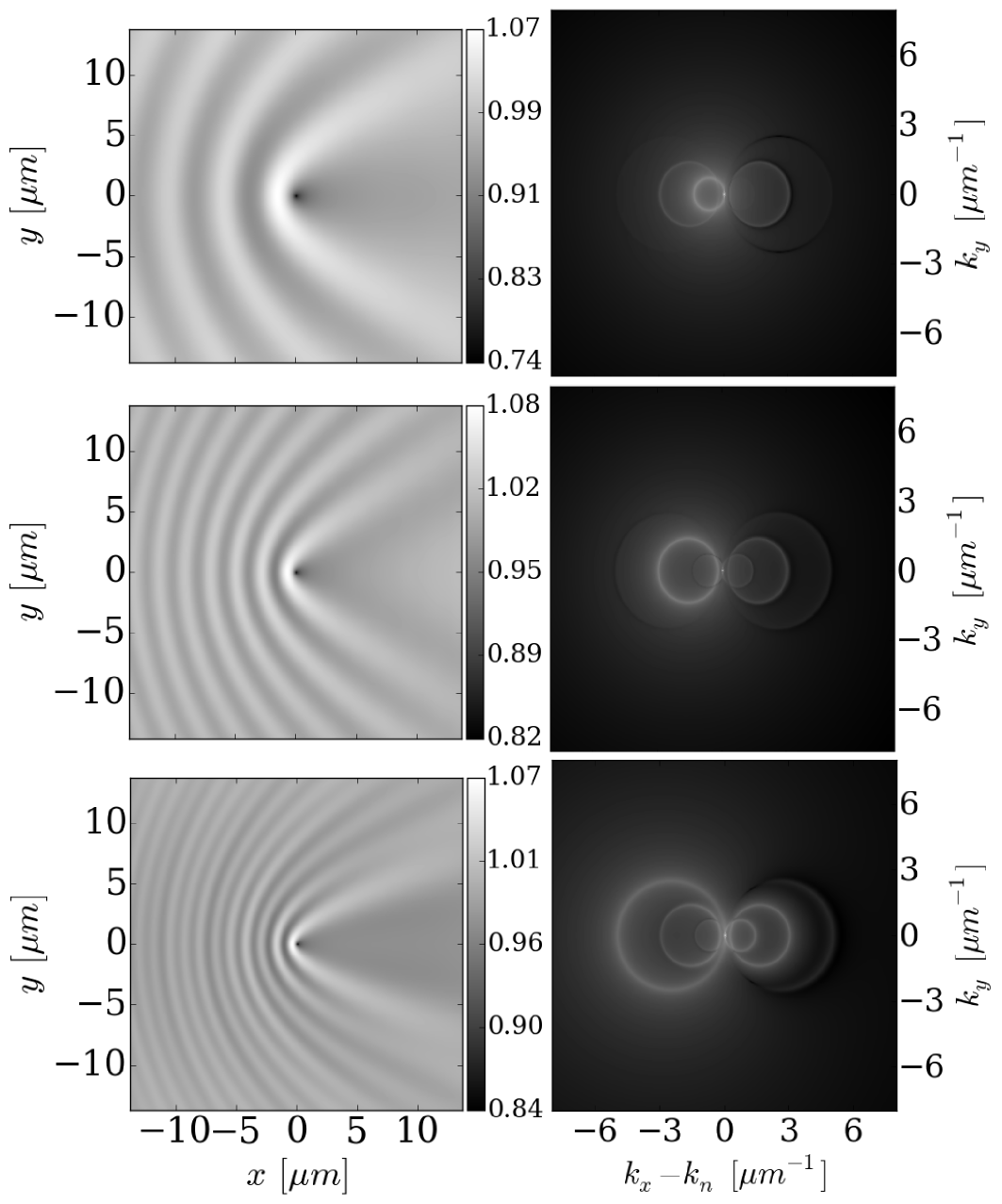}
\caption{(Color online) OPO for finite and positive signal momentum in
  the linear response approximation for homogeneous pumping. Rescaled
  OPO emission within the linear response approximation in real $|\psi
  (\vect{r},\omega_n)|^2/|\psi_n|^2$ (left panels in linear scale) and
  momentum space $|\psi_{\tilde{\vect{k}}} (\omega_n)|^2$ (right
  panels in linear scale) filtered at the energies of the three OPO
  states: signal (top panels), pump (middle), and idler (bottom). The
  parameters are the same as those used in Fig.~2 of the main
  manuscript, with the exception of the signal and idler momenta,
  which are now at $k_s = 0.7$~$\mu$m$^{-1}$ and $k_i =
  2.5$~$\mu$m$^{-1}$ respectively. Each scale of variation for the
  profiles is plotted in the color-box next to the left panels. A cut
  in the $y=0$ direction for the three profiles is plotted in the
  bottom panel of S.~Fig.~\ref{fig:range}.}
\label{fig:ksp07}
\end{figure}
\begin{figure}[h!]
\centering
\includegraphics[width=1.0\columnwidth]{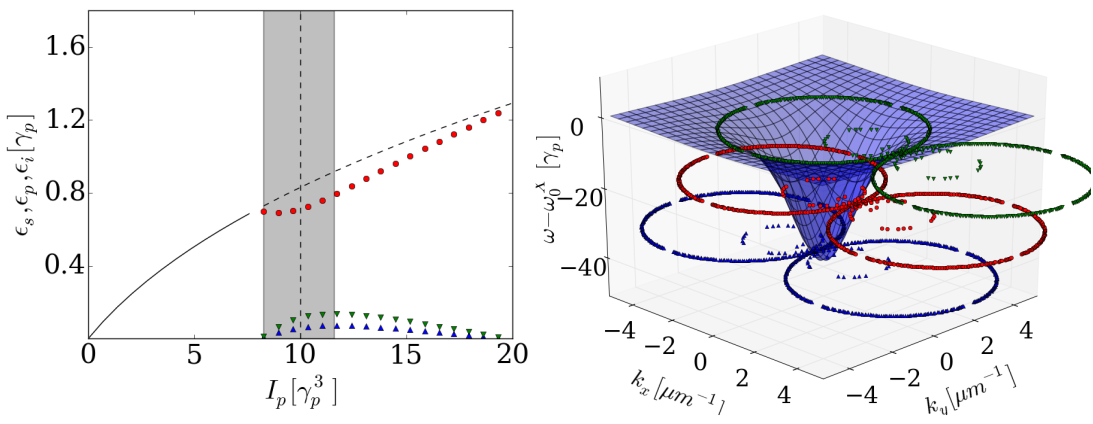}
\includegraphics[width=0.85\columnwidth]{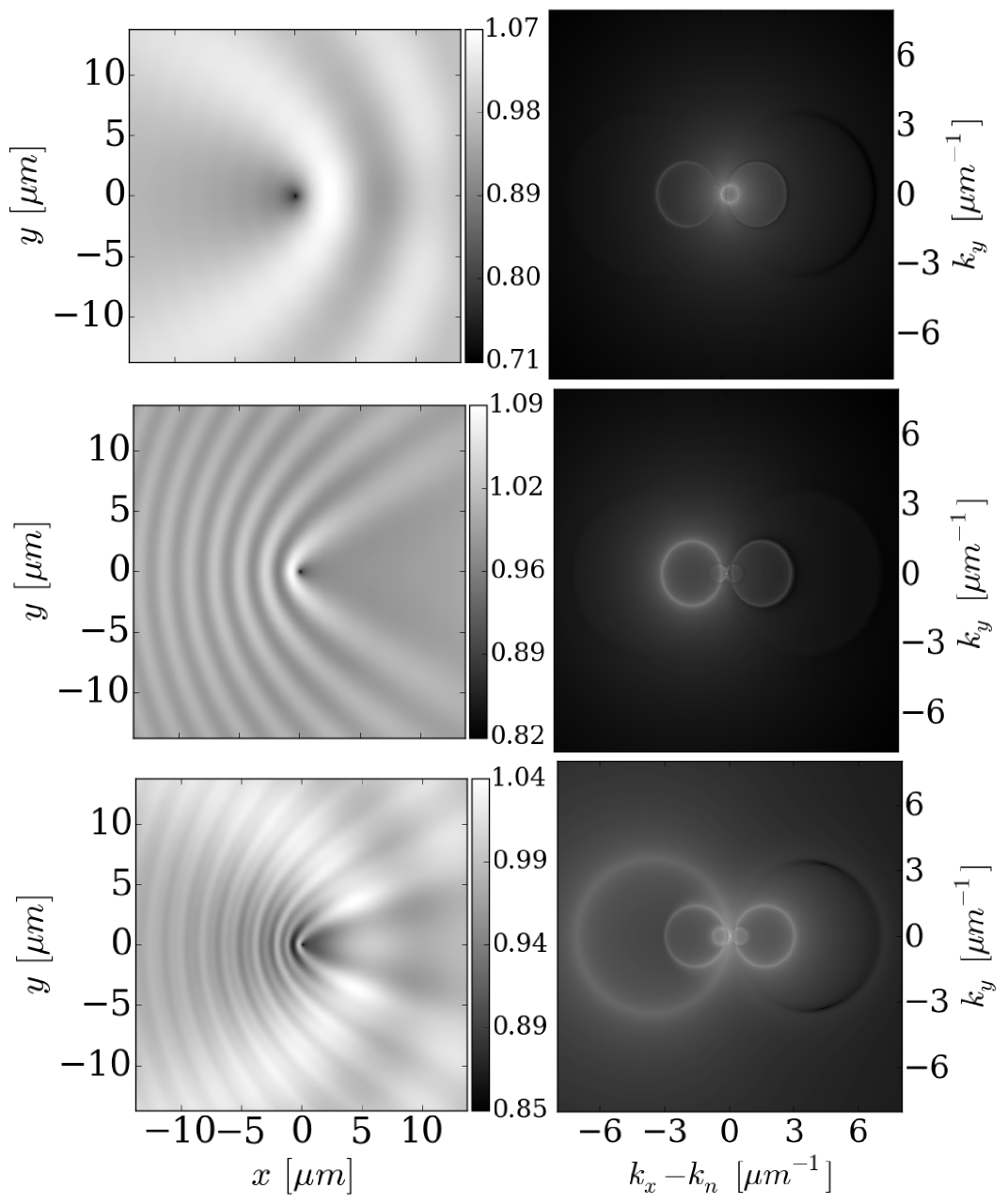}
\caption{(Color online) OPO for finite and negative signal momentum in
  the linear response approximation for homogeneous pumping. Rescaled
  OPO emission within the linear response approximation in real $|\psi
  (\vect{r},\omega_n)|^2/|\psi_n|^2$ (left panels in linear scale) and
  momentum space $|\psi_{\tilde{\vect{k}}} (\omega_n)|^2$ (right
  panels in linear scale) filtered at the energies of the three OPO
  states: signal (top panels), pump (middle), and idler (bottom). The
  parameters are the same as those used in Fig.~2 of the main
  manuscript, with the exception of the signal and idler momenta,
  which are now at $k_s = -0.4$~$\mu$m$^{-1}$ and $k_i =
  3.6$~$\mu$m$^{-1}$ respectively. A cut in the $y=0$ direction for
  the three profiles is plotted in the top panel of
  S.~Fig.~\ref{fig:range}.}
\label{fig:ksm04}
\end{figure}
\begin{figure}[h!]
\includegraphics[width=0.8\columnwidth]{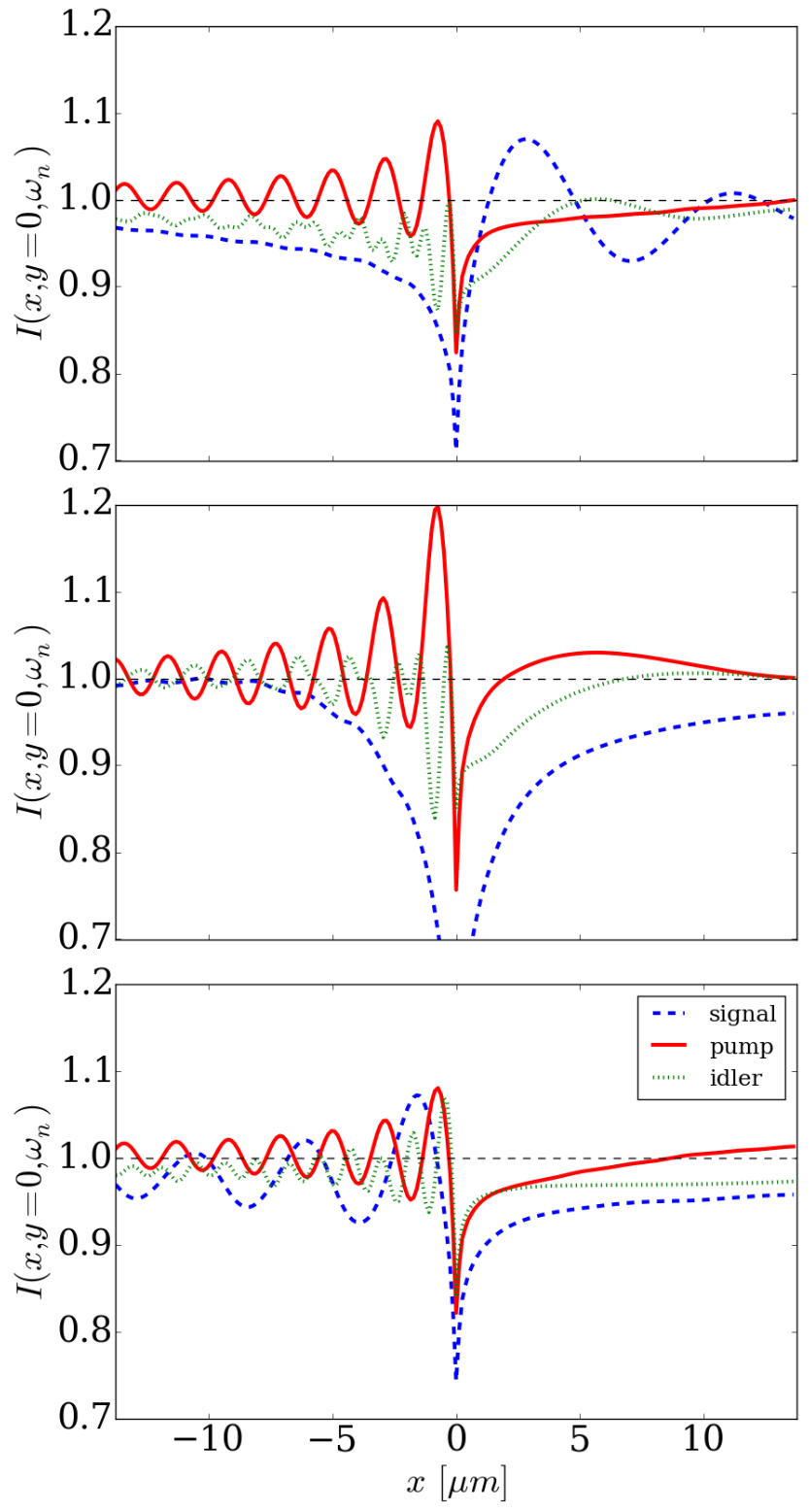}
\caption{(Color online) Real space signal, pump and idler OPO
  one-dimensional filtered profiles in the linear response
  approximation. OPO filtered emissions along the $y=0$ direction, $I
  (x, y=0, \omega_n) = |\psi(x,y=0, \omega_n)|^2/|\psi_n|^2$ rescaled
  by the mean-field solution $\psi_n$ in absence of the defect, for
  the three cases analysed above within the linear response
  approximation: the case of a signal at $k_s = -0.4$~$\mu$m$^{-1}$
  (top panel) corresponds to the same conditions as
  S.~Fig.~\ref{fig:ksm04} , a signal at $k_s = 0.0$~$\mu$m$^{-1}$
  (middle panel) corresponds to Fig.~2 of the main manuscript, and a
  signal at $k_s = 0.7$~$\mu$m$^{-1}$ (bottom panel) corresponds to
  S.~Fig.~\ref{fig:ksp07}. In each panel we plot, on the same linear
  scale, the filtered profiles of signal ([blue] dashed line), pump
  ([red] solid), and idler ([green] dotted), while the horizontal gray
  dashed lines represent the values of the mean-field emission prior
  to adding a defect (rescaled here to $1$).}
\label{fig:range}
\end{figure}
\begin{figure}[h!]
\includegraphics[width=1.0\columnwidth]{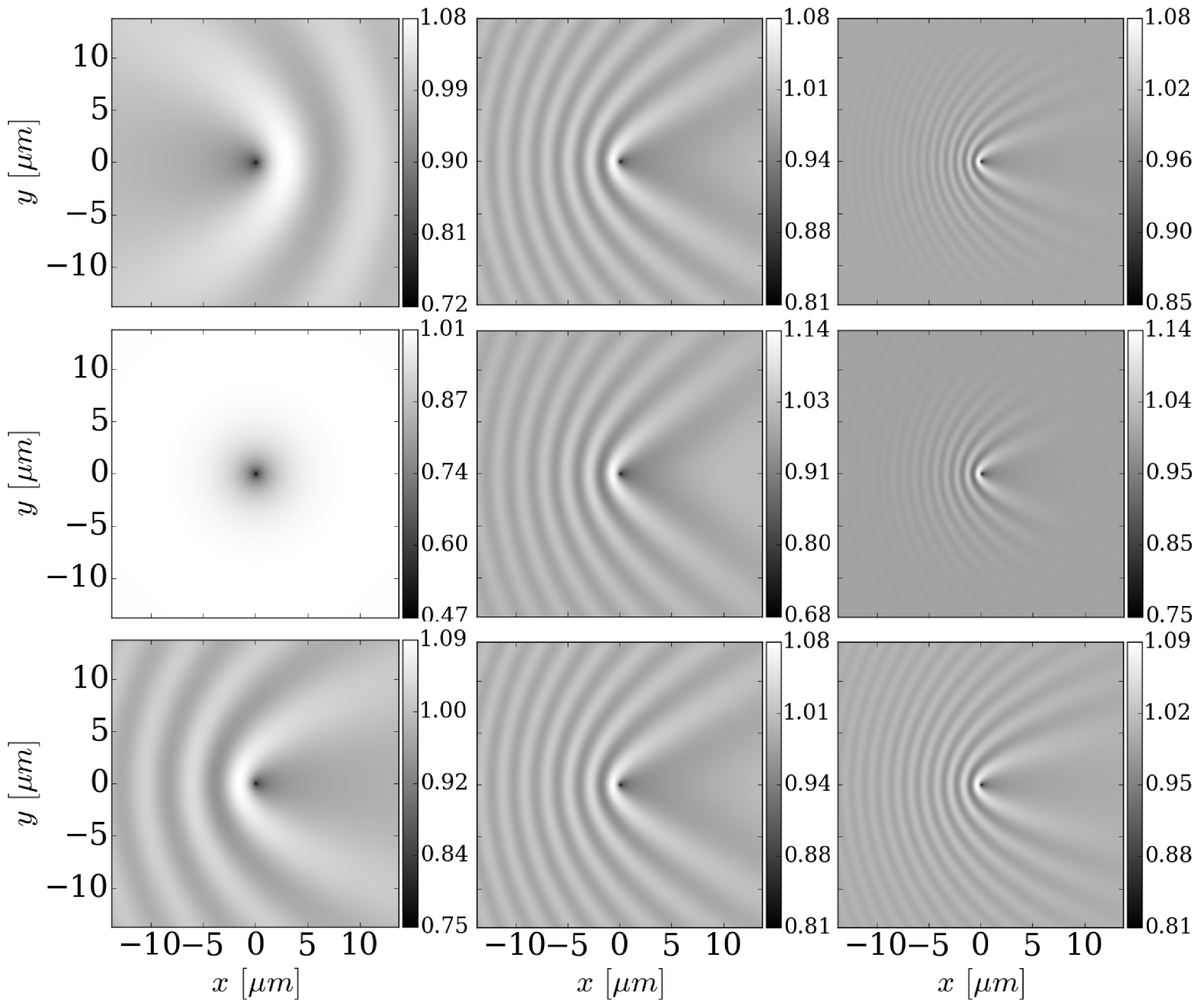}
\caption{Real space profiles of three uncoupled signal, pump and idler
  fluids. The rescaled profiles $|\psi
  (\vect{r},\omega_n)|^2/|\psi_n|^2$ are obtained by setting all
  off-diagonal couplings in Eq.~(6) of the main manuscript to zero,
  resulting in three uncoupled signal (left column), pump (middle
  column), and idler (right column) fluids. The three rows correspond
  to the three different cases previously analysed within the linear
  response approximation: the case of a signal at $k_s =
  -0.4$~$\mu$m$^{-1}$ (top row) corresponds to the same conditions as
  S.~Fig.~\ref{fig:ksm04} , a signal at $k_s = 0.0$~$\mu$m$^{-1}$
  (middle row) corresponds to Fig.~2 of the main manuscript, and a
  signal at $k_s = 0.7$~$\mu$m$^{-1}$ (bottom row) corresponds to
  S.~Fig.~\ref{fig:ksp07}.}
\label{fig:uncou}
\end{figure}
%
\section{Linear response}
\label{sec:analy}
In the manuscript we also make use of the linear response
approximation to analyse the OPO response to a static defect, valid
for a homogeneous pump scheme. In order to do so, we first evaluate
the Bogoliubov matrix given in Eq.~(6) of the manuscript, whose
eigenvalues determine the spectrum of collective excitations. We plot
in S.~Fig.~\ref{fig:bogol} a typical collective dispersion (here we
consider the same system parameters as the ones used in Fig.~2 of the
manuscript), by plotting the real part of the Bogoliubov matrix
eigenvalues $\Re \omega_{n,(u,v),\vect{k}-\vect{k}_n}$ as a function
of $k_x - k_n$ (cut at $k_y=0$). Note that the Rayleigh rings can be
found by finding the intersections $\Re
\omega_{n,(u,v),\vect{k}-\vect{k}_n}=0$.

As also done for the full numerics, we
consider there a $\delta$-like disorder potential $V_d(\vect{r}) = g_d
\delta(\vect{r} - \vect{r}_0)$, and Figs.~1 and~2 of the manuscript,
as well as S.~Figs.~\ref{fig:ksp07} and~\ref{fig:ksm04} below, are
plotted for this case.
We have however checked that our results do not depend on the specific
shape of the defect potential: In particular we have also considered
the response to defects with smoothen Gaussian-like profiles, whose
effect is only to partially weaken the upstream modulations emission
in real space.

We have seen in the manuscript that both experimentally as well as in
the full numerical analysis, one obtains OPO conditions where the
signal momentum is very close to zero, $\vect{k}_s \sim 0$. The reason
why OPO selects almost zero momentum signal conditions is still
awaiting a theoretical explaination.
In contrast, the linear response approximation allows to access very
different OPO mean-field conditions, for which, by leaving unaltered
the value of the pump momentum, the signal can appear at finite
momentum values, either positive or negative.
This is a peculiarity of this approximation scheme, where one can show
that, at mean-field level, one has the possibility of choosing
different values of the signal momentum $\vect{k}_s$ (and thus the one
of the idler $\vect{k}_i$), and that this choice range is quite broad
particularly when the pump power is close to the lower pump threshold
for OPO --- the stability region for OPO is plotted as a shaded grey
region in the top left panels of S.~Figs.~\ref{fig:ksp07}
and~\ref{fig:ksm04}. This is a well known ``selection problem'' for
parametric scattering (see Ref.~\cite{wouters07:prb}): The reason why
in full numerical calculations, parametric scattering processes select
a signal with a momentum close to zero, already very close to the
lower pump thresold for OPO, is still awaiting for an explaination. In
particular, this quest cannot be addressed within a spatially
homogeneous approximation where the three mean-field solutions for
pump, signal, and idler states can be describes by plane waves.
One can only show that, within the same mean-field approximation
scheme, when increasing the pump power towards the upper threshold for
OPO, the blue-shift of the LP polariton dispersion due to the
increasing mean-field polariton density, causes the signal momentum to
converge towards zero~\cite{whittaker05} $\vect{k}_s \to 0$.

\paragraph{OPO conditions with a finite momentum signal ---}
Given the freedom of choice for the signal momentum $\vect{k}_s$ close
to the lower OPO threshold, we consider here two additional cases,
that could not be studied neither experimentally, nor within a full
numerical approach, but that instead we can easily analyse within the
linear response theory.
In particular, we have left fixed the pumping conditions
$k_p=1.6$~$\mu$m$^{-1}$ and $\omega_p-\omega_X^0=-1.25$~meV and
considered two opposite situations.

In the first case, the signal has a finite and positive momentum $k_s
= 0.7$~$\mu$m$^{-1}$ and thus the idler is at low momentum, $k_i
=2.5$~$\mu$m$^{-1}$. The results are shown in S.~Fig.~\ref{fig:ksp07}.
Here we see that all six Rayleigh rings are clearly visible and, in
addition, as the idler is at lower momentum compared to the case
considered in Figs.~1 and~2 of the manuscript, and thus its dispersion
steeper, the idler group velocity is large enough to appreciate the
modulation of the Rayleigh ring associated to this state. As a result,
each of the three filtered OPO emissions exhibits as the strongest
modulation the one coming from its own Rayleigh ring, included the
signal which is now at finite and large momentum. In this case, the
OPO response of each filtered state profile looks completely
independent from the other, as if we were pumping each state
independently.

In the second case shown in S.~Fig.~\ref{fig:ksm04} , the signal is
finite and negative, $k_s = -0.4$~$\mu$m$^{-1}$, and the idler is now
at very large momentum, $k_i = 3.6$~$\mu$m$^{-1}$, where its
dispersion is very exciton-like and flat, and thus the idler has a
very small group velocity and its own modulations are visible only
very close to the defect. For this case, we can appreciate in the
idler filtered profile overlapped modulations from all the three state
Rayleigh rings (note that because the signal is at negative momentum,
its modulations have an opposite direction compared to the ones of
pump and idler), while in the signal we can mostly see the signal long
wavelength modulations and only very weakly the pump one.

We can compare the different modulation strengths of the three OPO
profiles by looking at the color bars plotted next to the profiles. In
order to better compare them on the same plot, we show in
S.~Fig.~\ref{fig:range} the one-dimensional OPO filtered emissions
along the $y=0$ direction for the three cases analysed above. While
for the OPO conditions with a signal at $k_s = 0.0$~$\mu$m$^{-1}$
(middle panel, corresponding to Fig.~2 of the main manuscript), the
imprinted modulations from the pump are hardly visible, and can only
be appreciated after a Gaussian filter manipulation, both OPO cases
with a finite momentum signal result in modulations in the signal with
an amplitude of the same order of magnitude of both pump and idler
fluids.

In S.~Fig.~\ref{fig:uncou} we instead show the real space profiles
obtained by setting all off-diagonal couplings in the Bogoliubov
matrix $\mathcal{L}_{\vect{k}}$ of Eq.~(6) of the main manuscript to
zero, resulting in three uncoupled signal (left column), pump (middle
column), and idler (right column) fluids. This underlines the
importance of the coupling between the three fluids in the three
different OPO regimes previously analysed within the linear response
approximation. In particular, for the OPO conditions such that the
signal has a finite and positive momentum $\vect{k}_s$ (bottom row of
S.~Fig.~\ref{fig:uncou}, which corresponds to the conditions shown in
S.~Fig.~\ref{fig:ksp07}), the coupling has little effect and the three
fluid respond to the defect in practice in a independent way (each
Rayleigh ring influences its own fluid). The OPO condition with a
finite and negative momentum $\vect{k}_s$, which corresponds to the
conditions shown in S.~Fig.~\ref{fig:ksm04}, are shown in the top row
of S.~Fig.~\ref{fig:uncou}: Here, we can see that the coupling between
the three fluids plays an effect. The modulations of the pump can be
appreciated both in the signal profile (though weakly), as well as in
the idler profile. In the idler fluid its own modulations can only
propagate very close to the defect and thus the only really
appreciable modulations are the ones inherited from the pump. Finally,
for the experimentally relevant case of an OPO with a signal at zero
momentum (middle row of S.~Fig.~\ref{fig:uncou}, which corresponds to
the conditions shown in Fig.~2 of the manucript), there is also a role
played by the coupling, though, as already thoroghly analysed, the
modulations in the signal inhereted from the pump can only be
appreciated after a Gaussian filtering manipulation.

Finally note that, as it also happens for the OPO conditions shown in
Fig.~1 of the manuscript, the subsonic to supersonic crossover of the
pump-only state~\cite{amo09_b} happens at pump intensities well above
the region of stability of OPO --- the shaded gray regions of
S.~Figs.~\ref{fig:ksp07} and~\ref{fig:ksm04} Thus, it is not possible
to study a case where the pump is already subsonic and at the same
time promotes stimulated scattering.


%

\end{document}